# A Probabilistic Attack on NP-complete Problems

Alexander Y. Davydov*

June 30, 2011

**Abstract**

Using the probability theory-based approach, this paper reveals the equivalence of an arbitrary NP-complete problem to a problem of checking whether a level set of a specifically constructed harmonic cost function (with all diagonal entries of its Hessian matrix equal to zero) intersects with a unit hypercube in many-dimensional Euclidean space. This connection suggests the possibility that methods of continuous mathematics can provide crucial insights into the most intriguing open questions in modern complexity theory.

## 1 Introduction

Non-deterministic polynomial time complete (NP-complete) problems are of considerable theoretical and practical interest and play a central role in the theory of computational complexity in modern computer science. Currently, more than three thousand vital computational tasks in operations research, machine learning, hardware design, software verification, computational biology, and other fields have been shown to be NP-complete. The 'completeness' designates the property that, if an efficient (= polynomial-time) algorithm for solving any *one* of NP-complete problems could be found, then we would immediately have (as a minimum) an efficient algorithm for *all* problems in this class [1-4]. Despite persistent efforts by many talented researchers throughout several decades, it is not currently known whether NP-complete problems can be efficiently solved. An unproven conjecture broadly spread among complexity theorists is that such polynomial-time algorithm cannot exist. It is also a general belief that either proof or disproof of this conjecture can only be obtained through development of some new mathematical techniques.

In the present paper, a novel approach to tackling NP-complete problems is proposed bringing a fresh perspective on the subject matter. More specifically, our approach takes the problem from the realm of *discrete* mathematics and reformulates it in the realm of *continuous* mathematics, where it is then treated using tools of mathematical analysis and probability theory. The main idea of the proposed method stems from recognizing that, owing to exponentially large solution space for whichever NP-complete problem, any prospective approach to solving it by examining solution candidates sequentially, one by one, is predestined to fail in yielding an efficient algorithm, regardless of how smart and sophisticated this approach is. The more promising approach would be to learn how to manipulate *all* the solution candidates *simultaneously* and avoid detailed examination of a specific candidate prematurely. At first sight, it appears like an impossible task. Surprisingly, this paper shows that it is attainable. The two key elements to success are (i) introduction of a new set of variables using probabilistic reasoning and (ii) smart choice of a cost function to be minimized which is expressed in terms of these variables.

* AlgoTerra LLC, 249 Rollins Avenue, Suite 202, Rockville, MD 20852, USA
E-mail: *alex.davydov@algoterra.com*

For definiteness, we will consider a specific NP-complete problem, the *3-bit Exact Cover* (*EC3*) which is equivalent to *Positive 1-in-3 SAT* [5]. The problem is to decide whether there exists an assignment of $N$ bits $Z=(z_1,z_2,\ldots,z_N)$, each taking value *0* or *1*, such that $M$ clauses (constraints) are simultaneously satisfied. Each clause involves exactly three bits, say $z_k, z_m$ and $z_n$ with $k,m,n \in \{1,2,\ldots,N\}$, and is satisfied if and only if one of the bits is *1* and the other two are *0*, i.e., $z_k + z_m + z_n = 1$. It is assumed that, within each clause, the indices $k$, $m$, and $n$ are all distinct.

The rest of the paper is organized as follows. In Section 2, we reformulate the *EC3* problem in the realm of continuous mathematics using a new set of variables and a cost function with some rather remarkable properties. Section 3 presents an iterative algorithm for solving *EC3* that candidly exploits these properties. The illustrative examples of the algorithm's performance for different problem sizes are given in Section 4. The empirical evidence of the surprising *universality* of variable flows as the algorithm proceeds at low clauses-to-variables ratios is presented in Section 5. Finally, in Section 6, we discuss open issues related to computational complexity of the presented algorithm and summarize the results.

## 2 Reformulation of the EC3 Problem

### *2.1 New variables and cost function*

There are $2^N$ possible assignments of bits $z_k$, so checking them sequentially until a solution is found would take (on average) exponential time. Instead, one would like to evaluate all candidate solutions simultaneously and manipulate the whole pool of them in a manner ensuring that promising candidate solutions get *enhanced attention* while the unpromising ones become *discouraged* in some sense. To reach this goal, we introduce a set of new variables for the *EC3* problem as follows. With each bit $z_i$, we associate a probability $x_i$ that value of $z_i$ is chosen as *0*:

$$x_i = \Pr\{z_i = 0\},\ 0 \le x_i \le 1,\ i = 1, 2, \ldots, N. \qquad (1)$$

It follows that

$$1 - x_i = \Pr\{z_i = 1\},\ 0 \le x_i \le 1,\ i = 1, 2, \ldots, N.$$

Vice versa, given some value of $x_i \in [0,1]$, the associated bit $z_i$ is selected to be *0* or *1* at random with probabilities $x_i$ and $1-x_i$, respectively. Let $X=(x_1,x_2,\ldots,x_N)$ denote an $N$-element vector representing the whole set of new variables. Consider two limiting cases. In first case, let all components of vector $X$ be either *0* or *1*. Then we have no ambiguity in choosing the corresponding bits $Z$, i.e., there is one-to-one correspondence between $X$ and $Z$. In another limiting case, all components of vector $X$ are distinct from either *0* or *1*. Then, for a fixed vector $X$, we can choose any of $2^N$ possible assignments of bits $Z$, although possibly with different weights. In particular, when $X = (\frac{1}{2}, \frac{1}{2}, \ldots, \frac{1}{2})$, each assignment of bits has the same weight (probability), no one assignment is chosen more often (on average) than the other.

Now we would like to construct a continuous *cost function F* for the *EC3* problem with domain coinciding with the hypercube Ω: $0 \le x_i \le 1$, ($i$ = 1,…, $N$). The highly desired property of the cost function is its dependence upon the structure of the problem in such a way that makes it 'feel', in some sense, where the search for the satisfying assignments should be conducted. As a first step in search for such a function, let us find the probability $P_i$ that a clause $C_i$ is *unsatisfied* given some fixed $X$. Table 1 lists all possible assignments of three distinct bits involved in the clause $C_i$ along with the corresponding probabilities.

**Table 1:** List of bit assignments and their probabilities for an arbitrary clause

| **Assignments of bits in the *i*-th clause** $C_i = z_{k_i} + z_{m_i} + z_{n_i}$ | **Clause status** | **Probability** |
|---|---|---|
| 0 0 0 | unsatisfied | $x_{k_i} \cdot x_{m_i} \cdot x_{n_i}$ |
| 0 0 1 | *satisfied* | $x_{k_i} \cdot x_{m_i} \cdot (1 - x_{n_i})$ |
| 0 1 0 | *satisfied* | $x_{k_i} \cdot (1 - x_{m_i}) \cdot x_{n_i}$ |
| 0 1 1 | unsatisfied | $x_{k_i} \cdot (1 - x_{m_i}) \cdot (1 - x_{n_i})$ |
| 1 0 0 | *satisfied* | $(1 - x_{k_i}) \cdot x_{m_i} \cdot x_{n_i}$ |
| 1 0 1 | unsatisfied | $(1 - x_{k_i}) \cdot x_{m_i} \cdot (1 - x_{n_i})$ |
| 1 1 0 | unsatisfied | $(1 - x_{k_i}) \cdot (1 - x_{m_i}) \cdot x_{n_i}$ |
| 1 1 1 | unsatisfied | $(1 - x_{k_i}) \cdot (1 - x_{m_i}) \cdot (1 - x_{n_i})$ |

The probability $P_i$ reads

$$\begin{aligned} P_i = 1 - x_{k_i} \cdot x_{m_i} \cdot (1 - x_{n_i}) - x_{k_i} \cdot (1 - x_{m_i}) \cdot x_{n_i} - (1 - x_{k_i}) \cdot x_{m_i} \cdot x_{n_i} = \\ 1 + 3 \cdot x_{k_i} \cdot x_{m_i} \cdot x_{n_i} - x_{k_i} \cdot x_{m_i} - x_{m_i} \cdot x_{n_i} - x_{k_i} \cdot x_{n_i} \end{aligned} \qquad (2)$$

Let us now define a *cost function* as a sum of probabilities $P_i$ for all clauses:

$$F(X) = \sum_{i=1}^{M} P_i(x_{k_i}, x_{m_i}, x_{n_i}) \qquad (3)$$

If all clauses were independent, then $F(X)$ would have a meaning of *probability* that at least one of the clauses of the corresponding *EC3* problem is unsatisfied by the assignment of bits chosen in accordance with $X$. However, clauses that contain common variables are dependent and so one cannot interpret $F$ as such probability in general case. Nevertheless, $F$ has a number of interesting properties which happen to be very useful for reformulating the *EC3* problem in terms of continuous mathematics. Let us now turn to investigating these properties.

### *2.2 Properties of the cost function F*

Let us start by noticing that the cost function $F$ contains all the information about the *EC3* problem: given $M$ clauses involving some or all $N$ variables, one can uniquely construct $F$, and, vice versa, if given $F$ in analytical form, one can easily recover the original problem in terms of clauses of logic variables. It is useful to expand the domain of $F$ to the *entire* $N$-dimensional space $\Re^N$. This allows us to focus on general behavior of the cost function without restrictions irrelevant to our goal and also has some other benefits as it will become evident shortly. The domain expansion is straightforward since $F$ is a non-singular function everywhere in $\Re^N$.

***Lemma 1.*** $F(X)$ is a *harmonic* function in $\Re^N$: $\nabla^2 F = 0$. Moreover, $F$ is a harmonic function of any non-empty subset of variables (assuming that values of the remaining variables are hold fixed).

<u>*Proof:*</u> Since each clause $C_i$ consists of three distinct bits, the corresponding probability $P_i$ depends on three different variables. From (2), we then observe that $\frac{\partial^2 P_i}{\partial x_{k_i}^2} = 0, \frac{\partial^2 P_i}{\partial x_{m_i}^2} = 0, \frac{\partial^2 P_i}{\partial x_{n_i}^2} = 0$ for any $i$, and the statement of the *Lemma 1* immediately follows. ■

From *Lemma 1* and the well-known property of harmonic functions in $\Re^N$, it follows that the cost function $F$ can attain its maximum and minimum values in an arbitrary compact domain $D$ only on the boundary $\delta D$ (*The maximum principle*). Notice also that the Hessian matrix of function $F$ has all its diagonal elements equal to zero everywhere in $\Re^N$.

***Lemma 2.*** $F(X) > 0$ everywhere in *interior* of the hypercube $\Omega$ ($0 < x_i < 1$, $i = 1,\ldots, N$).

<u>*Proof:*</u> By definition, $F$ is a sum of $P_i$ that all attain *non-negative* values inside the hypercube $\Omega$ because they have a meaning of probability there. Note that "inside the hypercube" means that all components of vector $X$ are different from either *0* or *1* although they might come very close to these limiting values. Recall that, any vector $X$ inside $\Omega$ corresponds to all $2^N$ assignments of logic variables although some of them are more probable that the other. Now suppose $F(X) = 0$ somewhere in the interior of $\Omega$. Then it follows that probabilities $P_i = 0$ for all $i = 1,\ldots, N$ simultaneously. This means that *all* clauses in the corresponding EC3 problem are satisfied with probability 1 *irrespective of the assignment of logic variables*. But clearly this is impossible. Thus, by *reductio ad absurdum*, $F(X)$ must be positive everywhere inside the hypercube $\Omega$. $F$ may or may not vanish only on the boundary $\partial\Omega$. ■

***Lemma 3.*** The spectrum of possible values of $F$ on the vertices of the hypercube $\Omega$ is discrete and consists of integers from 0 to $M$:

$$F = \{0, 1, 2, \ldots, M\} \text{ for } X \in \textit{Vertices of } \Omega$$

The value of $F$ at any particular vertex gives the number of unsatisfied clauses of the corresponding *EC3* problem, provided that logic variables are chosen in accordance with vector $X$ at this vertex.

<u>*Proof:*</u> As mentioned earlier, there is one-to-one correspondence between $Z$ and $X$ on the vertices of the hypercube $\Omega$. So there is no ambiguity in values of logic variables and hence every clause can be either satisfied or unsatisfied by the definite set of logic variables $Z$. This means that each $P_i$ is either *1* (clause is unsatisfied) or *0* (clause is satisfied). Since $F$ is a sum of all $P_i$, its spectrum of possible values consists of numbers of *unsatisfied* clauses and can vary from 0 to $M$. For instance, $F = M$ for $X = (0,0,\ldots,0)$ or $X = (1,1,\ldots,1)$. ■

***Lemma 4.*** If a level set defined by the equation $F(X) = 1$ passes through the *interior* of $\Omega$, then there exists at least one satisfying bit assignment for the *EC3* problem.

<u>*Proof:*</u> Recall that, inside the hypercube $\Omega$, each $P_i$ has the meaning of probability of an event $A_i$ that the corresponding clause $C_i$ is unsatisfied, i.e., $P_i = \Pr\{A_i(X)\},\ X \in \Omega$. From the *union bound* (or *Boole's inequality*) [6], it follows

$$\Pr\left\{\bigcup_{i=1}^{M} A_i\right\} \le \sum_{i=1}^{M} P_i \ , \tag{4}$$

which can be also recast as

$$1 - \Pr\left\{\overline{\bigcup_{i=1}^{M} A_i}\right\} \le \sum_{i=1}^{M} P_i \tag{4a}$$

Applying DeMorgan's law [7] to the lhs of (4a) and making use of the definition (3), one obtains

$$\Pr\left\{\bigcap_{i=1}^{M} \overline{A}_i\right\} \ge 1 - F(X) \tag{5}$$

If $F(X) = 1$ inside $\Omega$, then there exists an open set $X^* \in \Omega$ where $F(X^*) < 1$ due to continuity of the cost function. It then follows from (5) that, for such $X^*$, the probability of all clauses being simultaneously satisfied is *positive*: $\Pr\left\{\bigcap_{i=1}^{M} \overline{A}_i(X^*)\right\} > 0$, which proves the *Lemma 4*. ■

## *2.3 Reformulation of EC3 problem in the realm of continuous mathematics*

In section 2.2, we presented a simple procedure of how, for a given *EC3* problem with *N* logic variables and *M* clauses, a *harmonic* function *F* of *N* continuous variables can be constructed in the analytic form. Function *F* defines a level set (hypersurface) $S_0$ by the equation: $F(X) = 0$. From *Lemma 2* it follows that there are only two possibilities:

1. $S_0$ intersects the hypercube $\Omega$ ($0 \le x_i \le 1$, $i = 1, 2, \ldots, N$) on the boundary $\partial\Omega$;
2. $S_0$ does not have any common points with the hypercube $\Omega$.

From the meaning of *F*, it is obvious that, in Case #1, the corresponding *EC3* problem is satisfiable (and the vertices of $\Omega$ that belong to the hypersurface $S_0$ uniquely determine the satisfying assignments of logic variables) while, in Case #2, the problem has no solutions. Thus *EC3* can be thought of as a problem of *intersection* between a level set $S_0$ of a harmonic cost function *F* and a unit hypercube in $\Re^N$.

Alternatively, based on the *Lemma 4*, we can present the *EC3* problem as a question of whether a level set $S_1$ defined by the equation $F(X) = 1$ *intersects* with the *interior* of the hypercube $\Omega$.

In any case, we observe that the *EC3*, initially formulated as a problem in the realm of discrete mathematics, can be mapped into a problem in the realm of continuous mathematics. The reformulated problem can be then attacked using the methods of mathematical analysis, differential geometry, and manifold topology. New interesting algorithms for solving NP-complete problems may emerge as a result of these attacks. The next section presents an example of such an algorithm.

## 3 Iterative Algorithm for Solving the *EC3* Problem

Possibly the simplest new algorithm for the *ES3* problem belongs to the *gradient descent* family and is described as follows:

<u>*Baby Steps Gradient Descent (BSGD) Algorithm*</u>

(1) Construct cost function $F(X)$ using Eqs (2) and (3);

(2) Select a starting point $X^{(0)}$ inside $\Omega$ and a constant step parameter $\eta > 0$;

(3) **Repeat**

(3.1) Update: $x_n^{(k+1)} = x_n^{(k)} - \eta \cdot \dfrac{\partial F(X^{(k)})}{\partial x_n^{(k)}}$ for $n = 1, 2, \ldots, N$;

(3.2) **if** $x_n^{(k+1)} > 1$ **then** $x_n^{(k+1)} = 1$;

(3.3) **if** $x_n^{(k+1)} < 0$ **then** $x_n^{(k+1)} = 0$;

**Until** stopping criterion $X^{(k+1)} = X^{(k)}$ is satisfied

(4) Compute: $F(X)$;

(5) **if** $F(X) = 0$ **then** *EC3* has a solution given by *Z* chosen in accordance with *X*;
**otherwise** report that a single run from starting point $X^{(0)}$ failed to find a solution.

Several remarks are in order. First, *Lemma 1* guarantees that the algorithm presented above never becomes trapped at some local minimum simply because of absence of local minima for a harmonic function *in interior* of a compact domain. The only stationary points that can be encountered during evolution of *X* are the *saddle points* where $\nabla F = 0$. The saddle points form a set of measure zero in $\Re^N$, thus making the probability of *X* passing through them negligible. Nevertheless, saddles play a crucial role by being bifurcation points that separate different evolution paths. They may cause two trajectories that originate at close starting points to end up far away from each other. The noticeable saddle point in the interior of the hypercube Ω is $X_s = (\frac{2}{3}, \frac{2}{3}, \ldots, \frac{2}{3})$, as can be easily verified by differentiation of $P_i$ with respect to any variable. Hence this point must not be used to start the algorithm with. However, any other point in the vicinity of $X_s$ can serve as a starting point (provided it is not a saddle). Second remark concerns consecutive runs of the algorithm, each time starting from a different point. If, during the latest run, the solution is found, we stop; otherwise we continue until maximum time designated to find a solution has been exhausted. With each unsuccessful run of the algorithm, the confidence that there is no solution increases. Whether it is possible or not to obtain a quantitative estimate of the probability that solution exists in terms of *N*, *M*, and the number of unsuccessful runs *R* is an open question which we will discuss briefly in Section 6. Third, notice that consecutive runs of the algorithm can be implemented in parallel. Once the cost function is generated, steps (2) through (4) of the *BSGD* can be carried out on *K* different processors simultaneously resulting in almost linear speedup, with each processor using its own starting point generated at random according to some protocol.

## 4 Illustrative Examples

Let us consider several examples illustrating the performance of the algorithm described in Section 3. All problem instances presented below were generated randomly, and the starting points for consecutive runs of the *BSGD* algorithm were chosen at random lying on the surface of an *N*-dimensional sphere centered at $(\frac{1}{2}, \ldots, \frac{1}{2})$ with the radius 0.05. *PM* denotes the '*problem matrix*' of size *3 × M* which consists of indices of variables involved in clauses columnwise. The step parameter $\eta$ was chosen to be equal to 0.005.

A. Small-size problem: *N* = 15; *M* = 8 (M/N ≅ 0.53)

$$PM = \begin{matrix} 3 & 4 & 11 & 2 & 7 & 5 & 1 & 6 \\ 6 & 7 & 14 & 6 & 12 & 13 & 6 & 9 \\ 15 & 13 & 15 & 12 & 15 & 15 & 8 & 10 \end{matrix}$$

The solution was found after the very first run:

Z = (1, 0, 1, 1, 1, 0, 0, 0, 0, 1, 0, 1, 0, 1, 0)

Figures 1 and 2 show the evolution of the cost function and the variables $x_n\,(n=1,\ldots,N)$, respectively, as the algorithm proceeds. Notice that some $x_n$ are changing in non-monotonic manner.

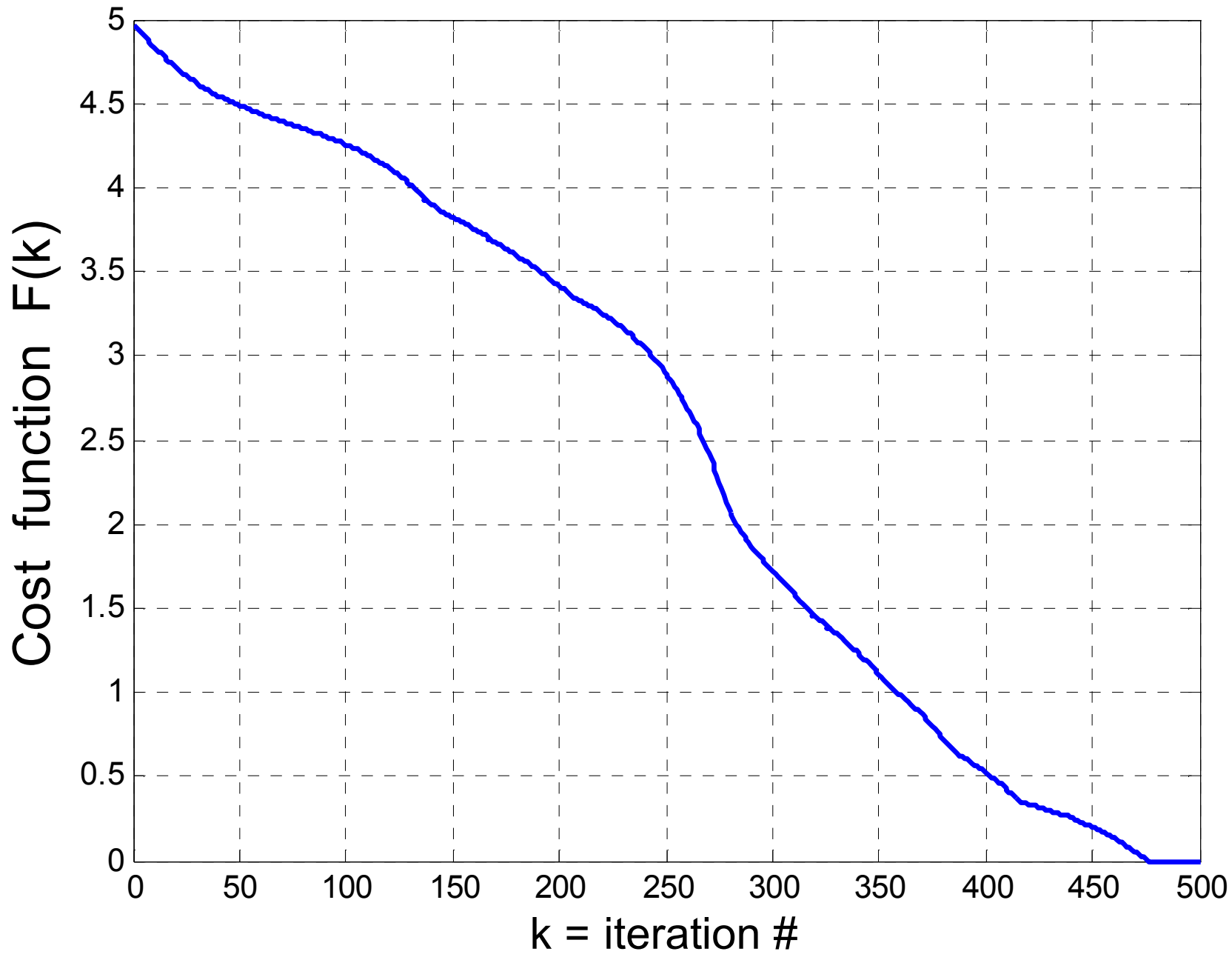

**Figure 1:** Cost function $F$ versus iteration number for the case $A$.

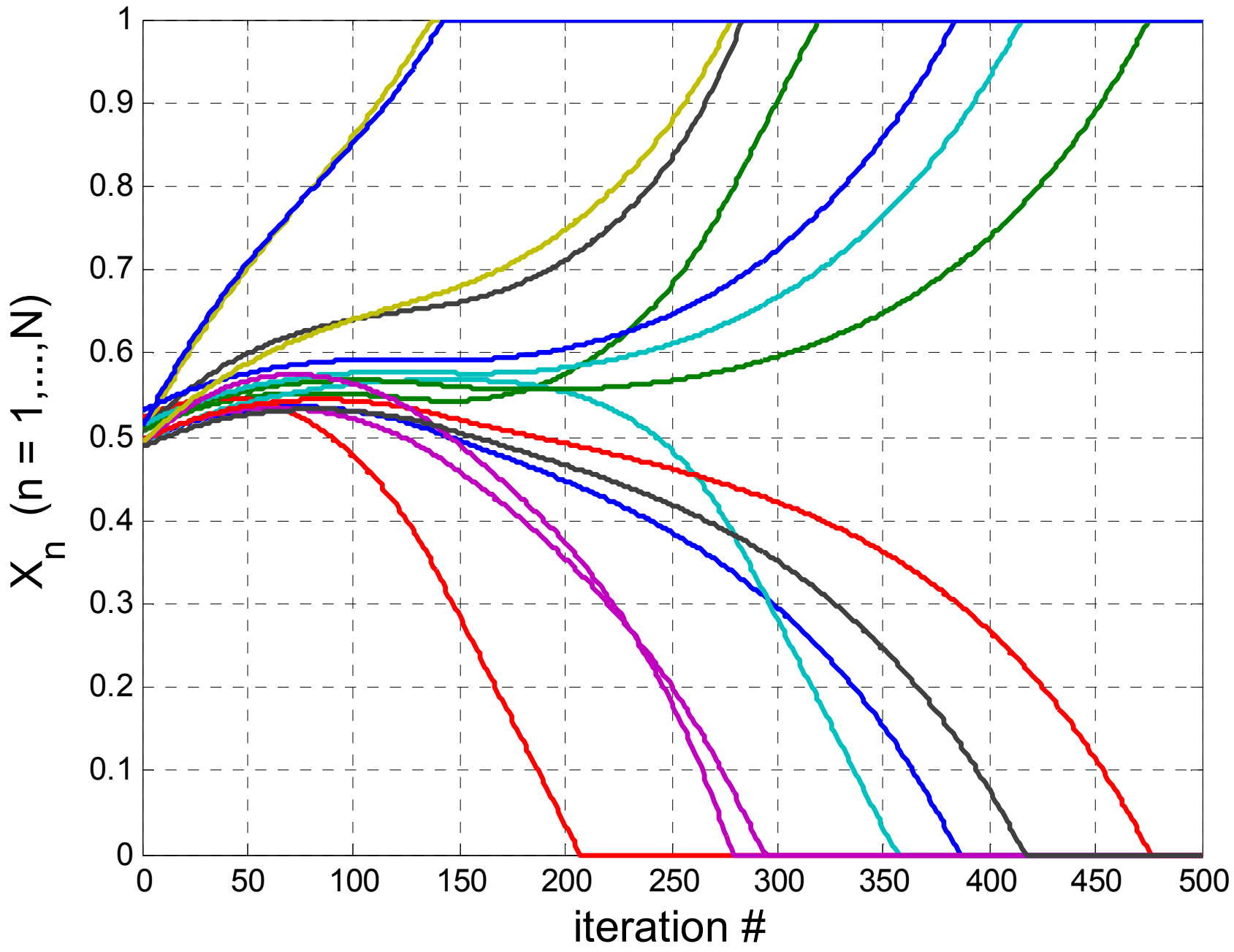

**Figure 2:** Evolution of variables $x_n$ for the case $A$.

B. Medium-size problem: $N = 100$; $M = 40$ (M/N = 0.4)

It took four successive runs of the *BSGD* algorithm to find a solution. The evolution curves for the cost function $F$ for all four runs are shown in Figure 3, and the evolution of the variables $x_n$ for the last successful run is presented in Figure 4.

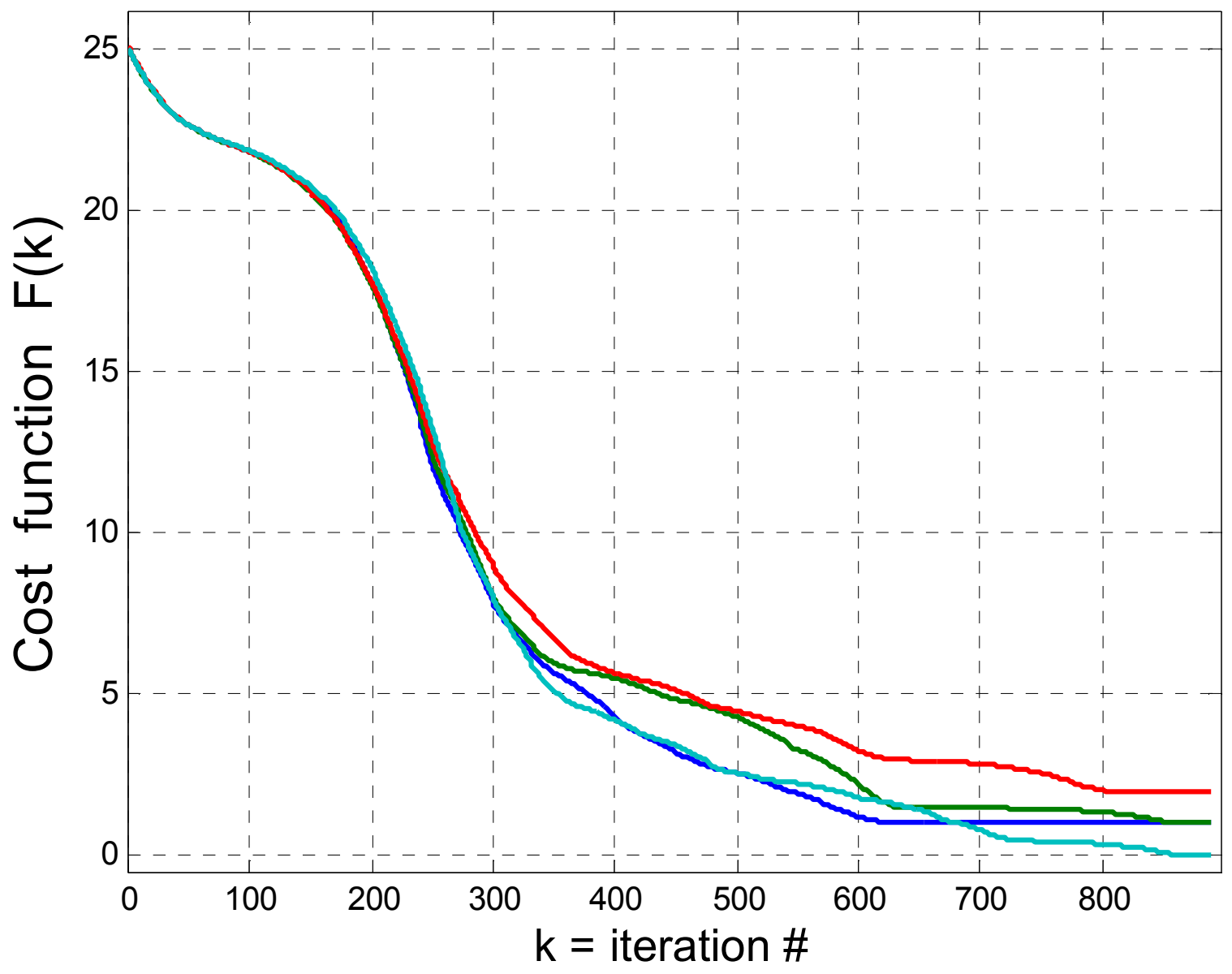

**Figure 3:** Cost function $F$ versus iteration number for the case $B$ (for four runs of the BSGD algorithm).

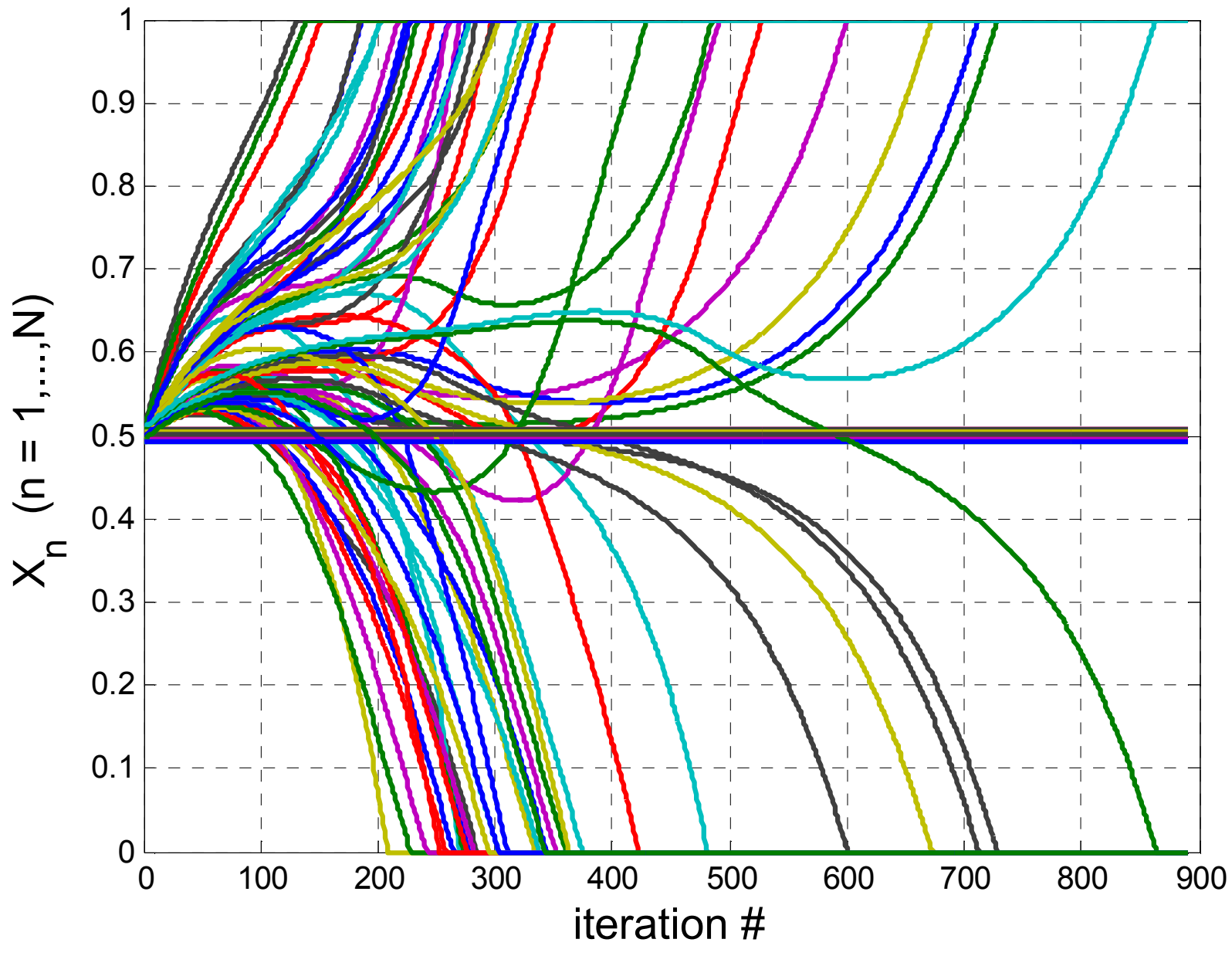

**Figure 4:** Evolution of variables $x_n$ in the successful run for the case $B$.

C. Large-size problem: $N = 1000$; $M = 250$ (M/N = 0.25)

It took three runs to find a solution. Figures 5 and 6 present the results (see captions for details).

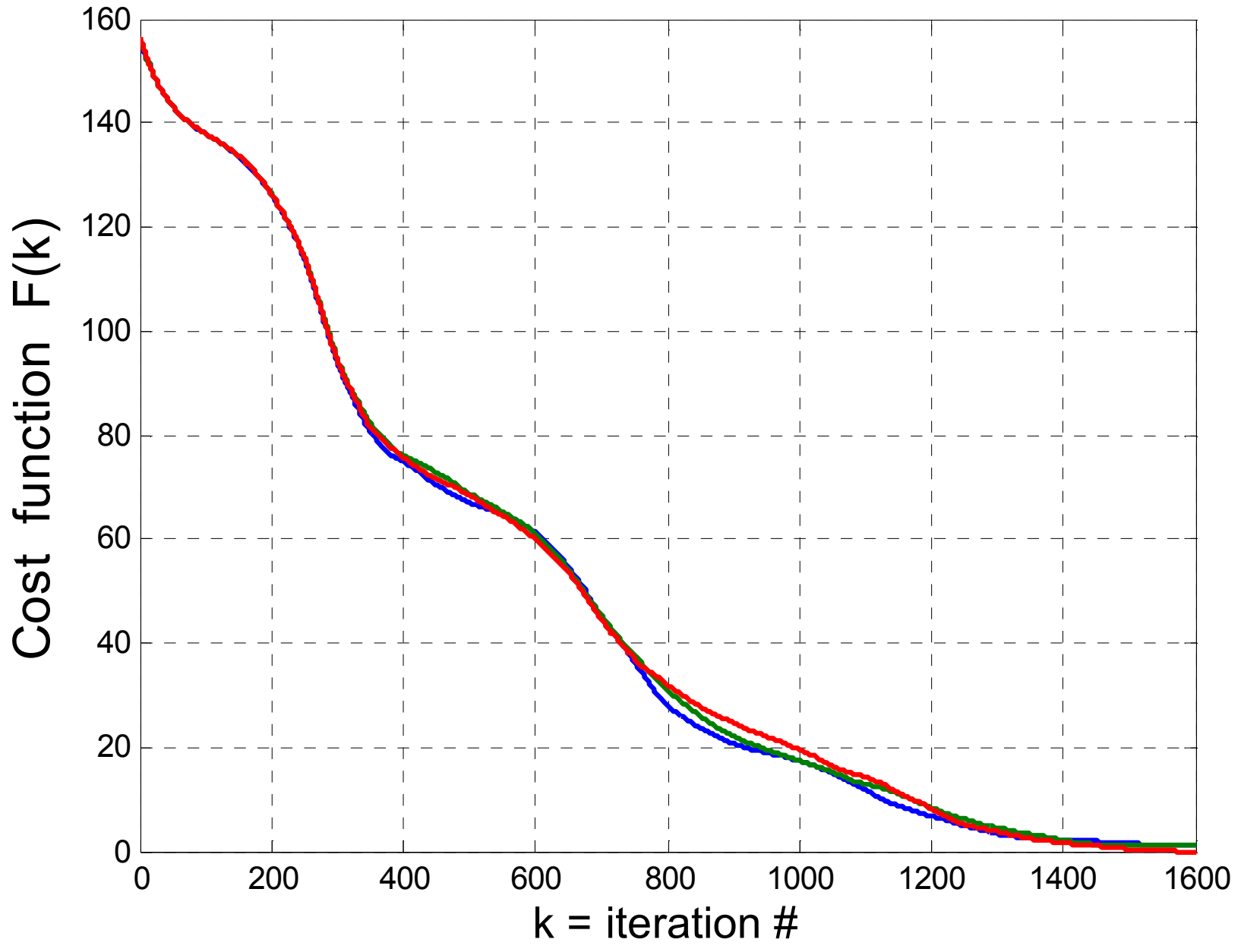

**Figure 5:** Cost function $F$ versus iteration number for the case $C$ (for three runs of the BSGD algorithm).

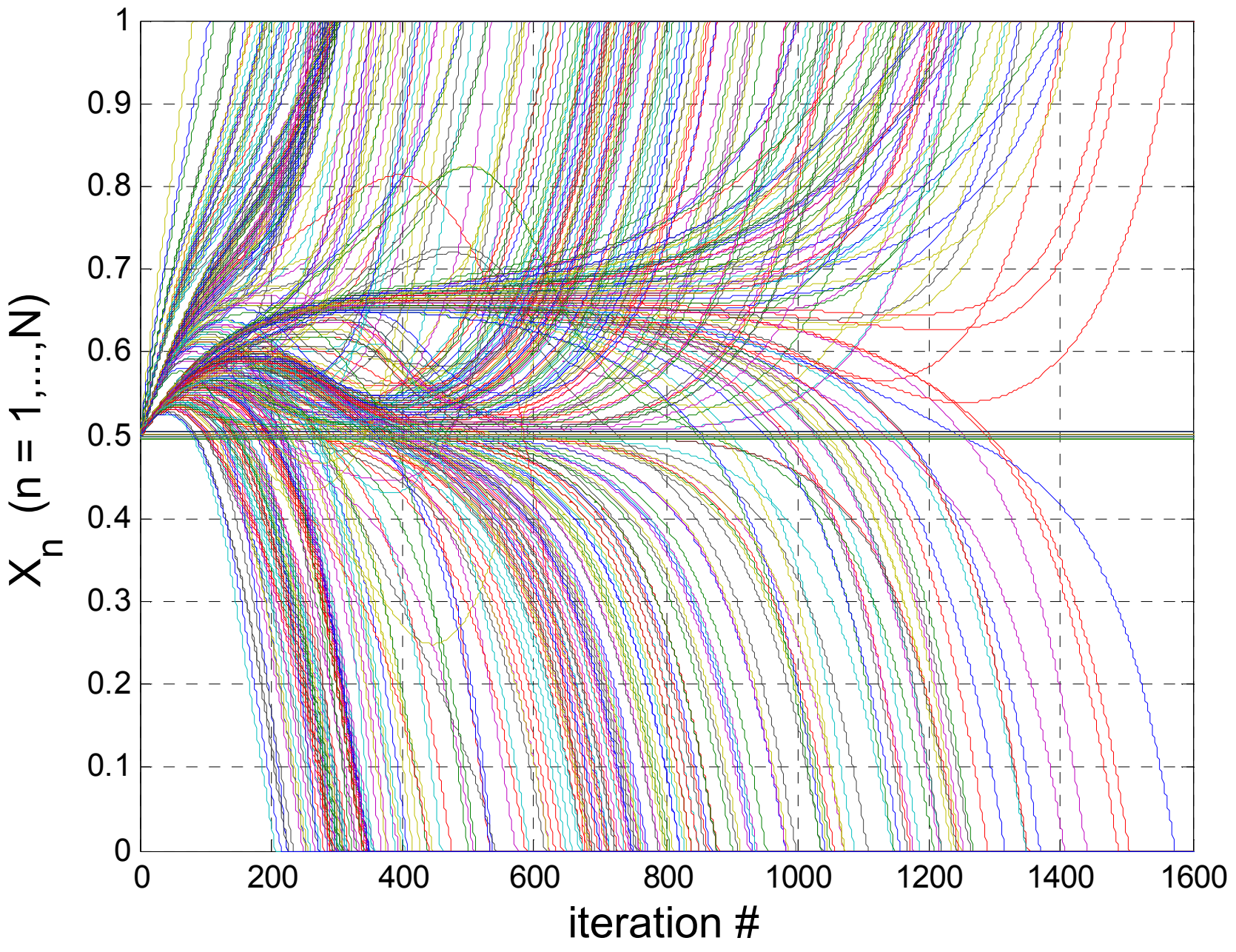

**Figure 6:** Evolution of variables $x_n$ in the successful run for the case $C$.

## 5 Universality of Variable Flows at Low Clauses-to-Variables Ratios

*- Happy families are all alike; every unhappy family is unhappy in its own way.*
*Leo Tolstoy, "Anna Karenina"*

There exists extensive empirical evidence suggesting that many constraint satisfaction problems exhibit phase transition phenomena from satisfiability to unsatisfiability as the ratio of the number of clauses (constraints) to the number of variables $r$ (= $M/N$) passes through some threshold value $r^*$ in the limit $N \to \infty$ [8-11]. For the *Positive 1-in-3 SAT* (or *EC3*) problem, it is found by numerical experiments that $r^* \approx 0.62$ [12]. Consequently, randomly generated instance of the *EC3* problem with large $N$ has a solution w.h.p. when $r << r^*$. Figure 7 shows the evolution of variables $x_n$ for the randomly generated *EC3* problem with $r = 0.025$ ($M$ = 25, $N$ = 1000). Surprisingly, these variable flows are *typical* and persist for different problem instances with small $r$ (which all have satisfying assignments w.h.p.). One can distinguish five families of well-separated variable flows, all originating from starting points around ½ (see Fig. 7):

**I** – Monotonically growing towards *1*;

**II** – Monotonically growing to a plateau at (roughly) *2/3* and then splitting into two sub-flows leading in the opposite directions to *1* and *0*, respectively;

**III** – Growing towards *1* first and then returning to a plateau at ½ where the flow splits into two sub-flows towards *1* and *0*, respectively;

**IV** – Growing towards *1* first and then changing the direction towards *0*;

**V** – Horizontal flow of irrelevant variables (those that are not involved in any clause).

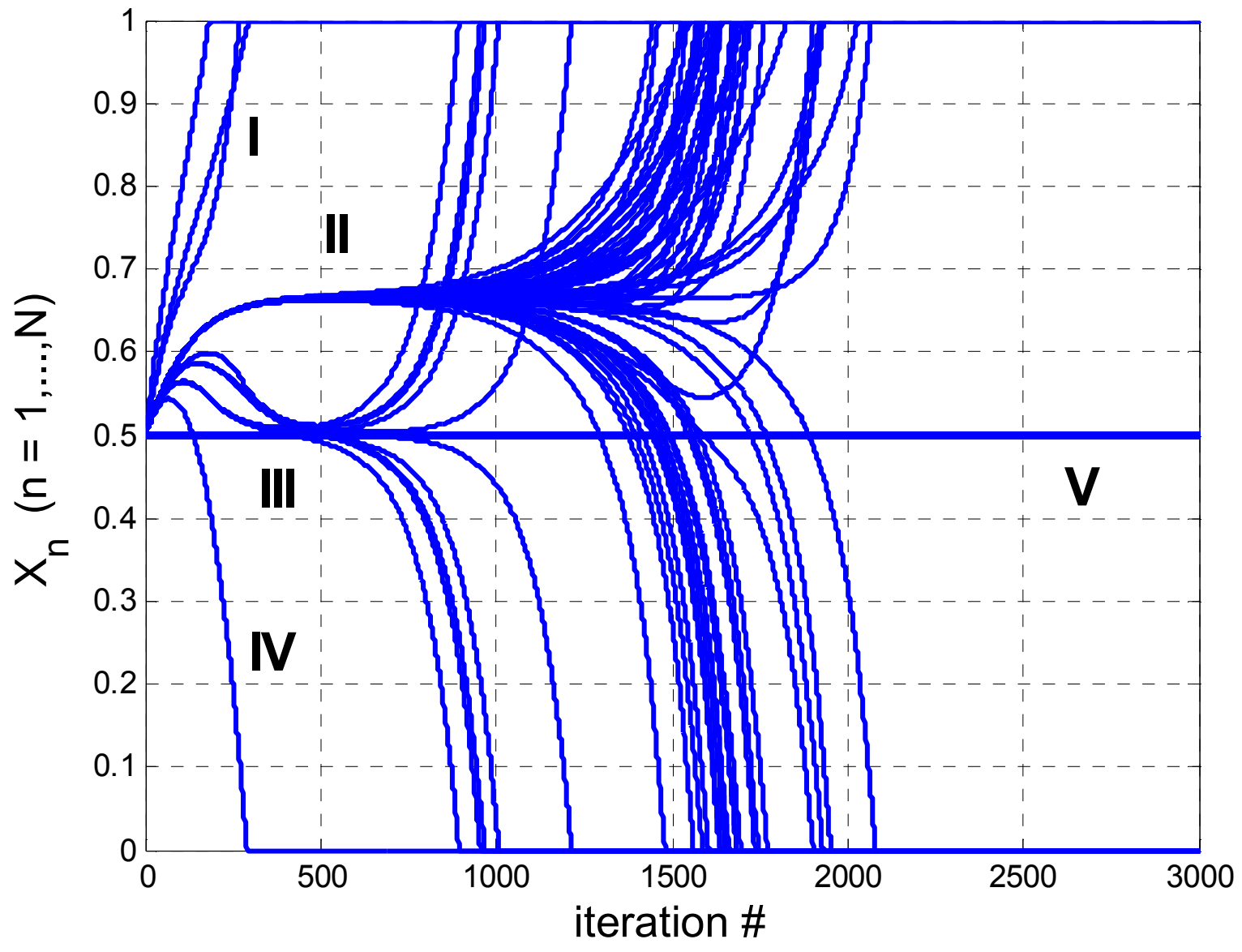

**Figure 7:** Typical variable flows for randomly generated *EC3* problem at low clauses-to-variables ratios (see text for details).

It is also worth noting that, for any variable and independently from the value of $r$, the slope of the corresponding trajectory can take only discrete set of values near the starting point close to ½. Specifically, for any $k \in \{1,2,\ldots,N\}$ it holds

$$x_k^{(2)} - x_k^{(1)} \cong \tfrac{1}{4}\eta \cdot C_k \,, \tag{6}$$

where $C_k \in \{0,1,2,\ldots,M\}$ denotes the total number of clauses that $x_k$ is involved in. Figure 8 clearly illustrates this fact by zooming in the leftmost region of Fig. 7.

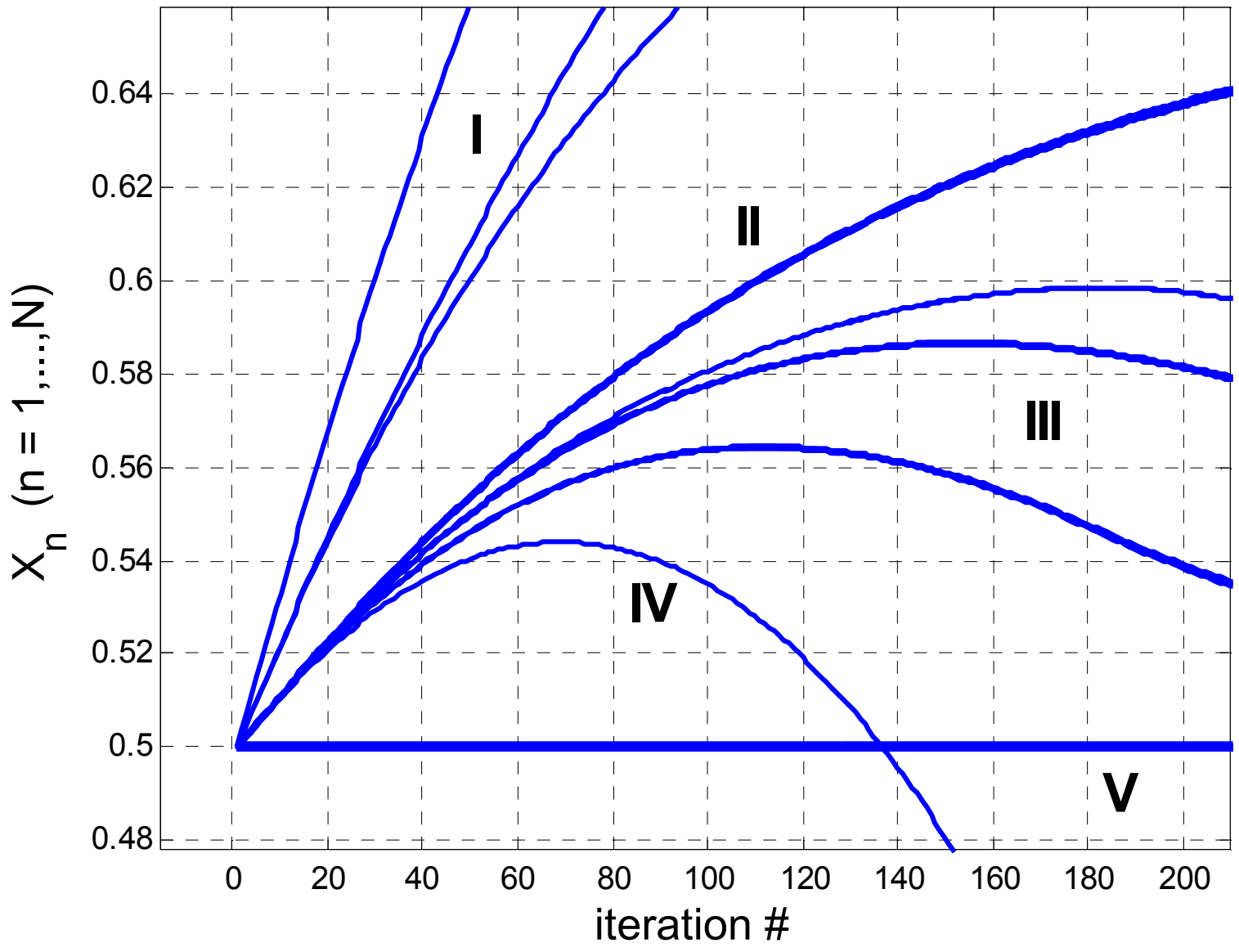

**Figure 8:** Zooming in the region of Fig. 7 near the starting points around ½; discreteness of the spectrum of starting slopes for variable trajectories can be clearly observed.

Numerical experiments show that, as the ratio $r$ increases, the highly ordered behavior of variable flows (as demonstrated in Figure 7) starts to crumble, with new trajectories not belonging to any of the abovementioned five families beginning to appear. With further increase of $r$, the basic five types of flow become wider and start to overlap while at the same time share of the 'irregular' trajectories grows. Finally, when ratio $r$ approaches the threshold value, only few variables follow the familiar paths, as illustrated in Table 2. The situation, to some extent, resembles the transition from laminar to turbulent flow in hydrodynamics, with control parameter $r$ playing the role of the Reynolds number. The nature of such behavior of variable flows for randomly generated *EC3* instances remains to be understood and maybe even quantified.

**Table 2:** Typical variable flows generated by the *BSGD* algorithm

| ***r*** | ***Solution found*** | ***No solutions found*** |
|---|---|---|
| 0.28 | 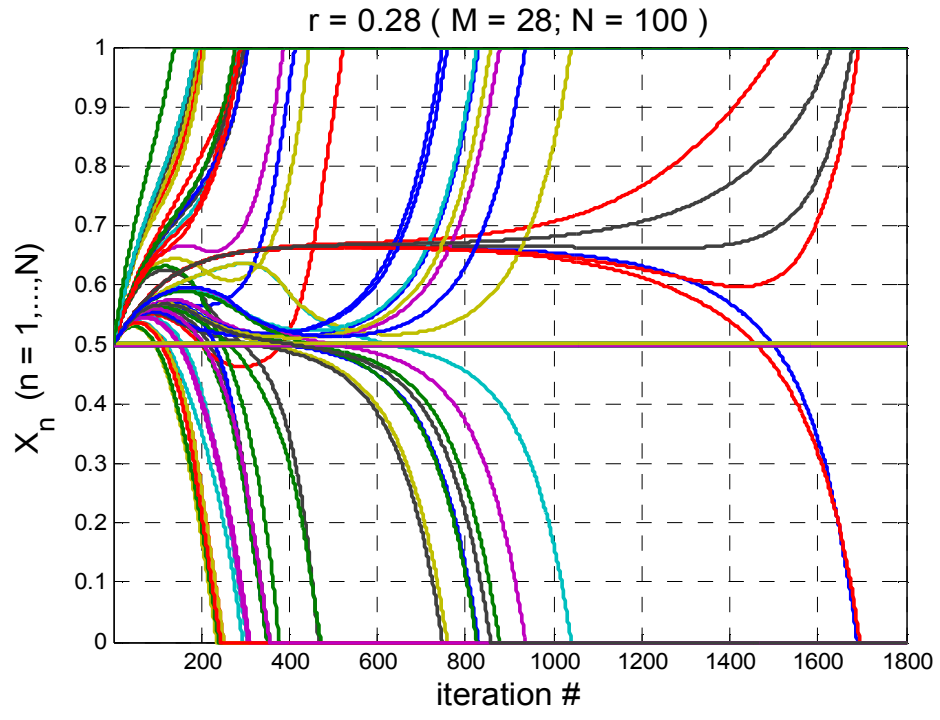 | • • • |
| 0.44 | 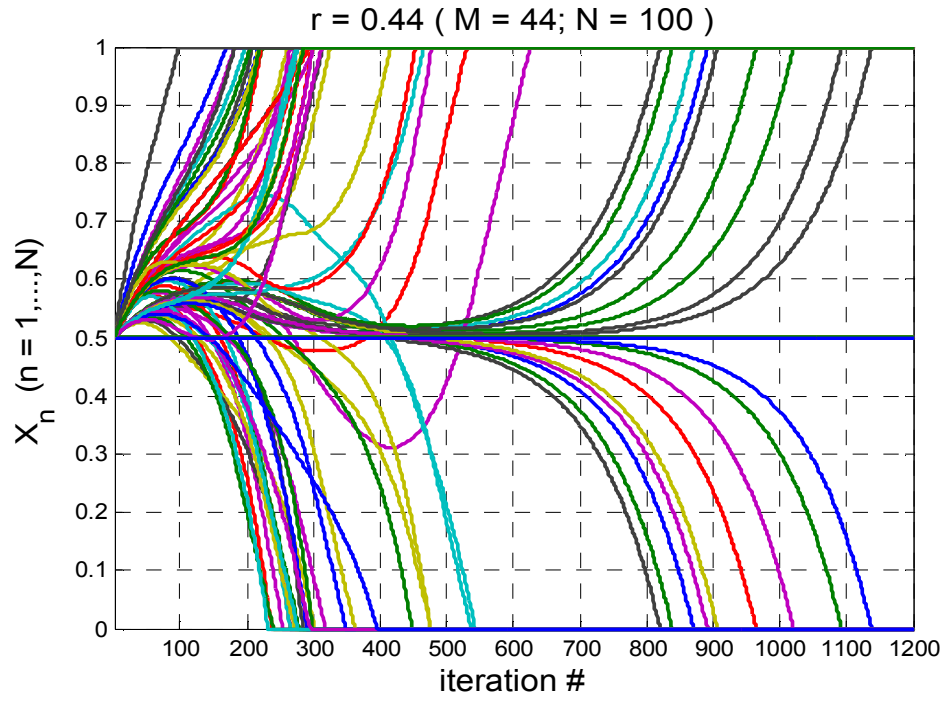 | 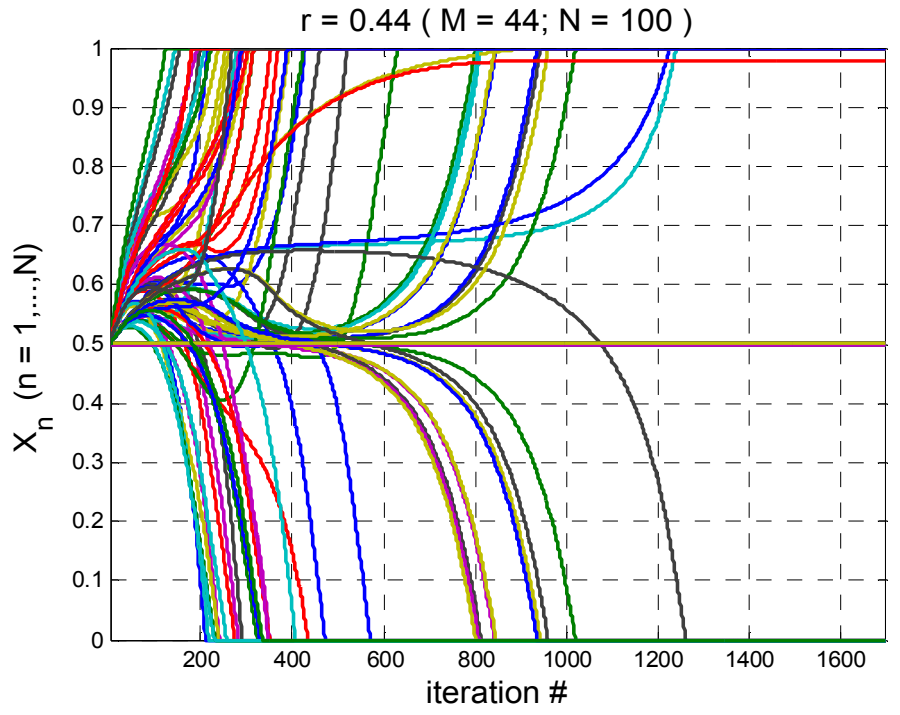 |
| 0.52 | 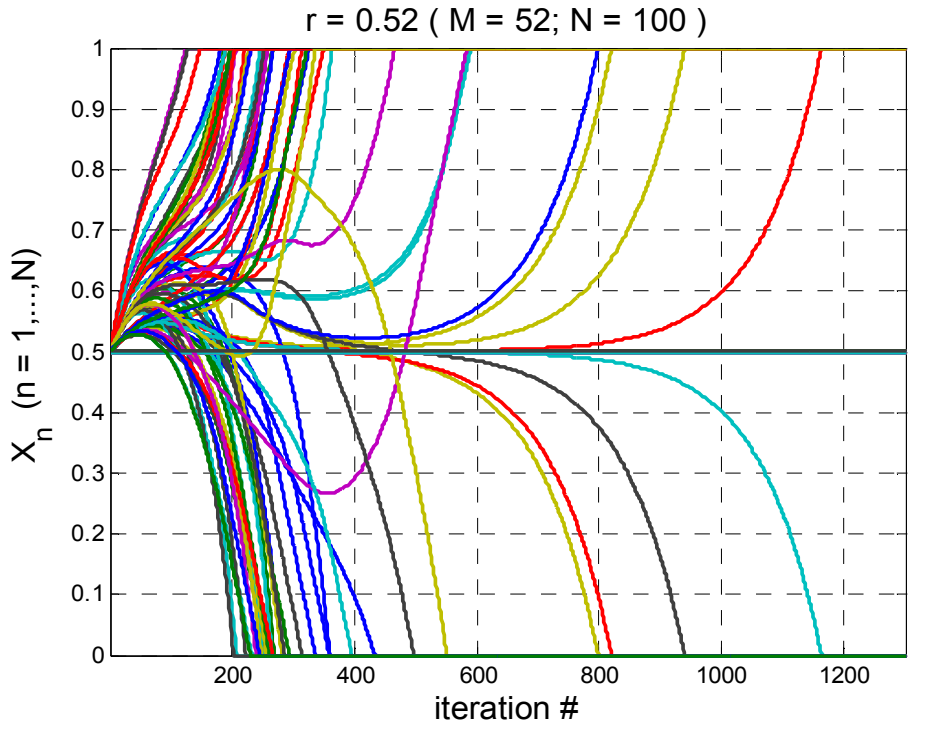 | 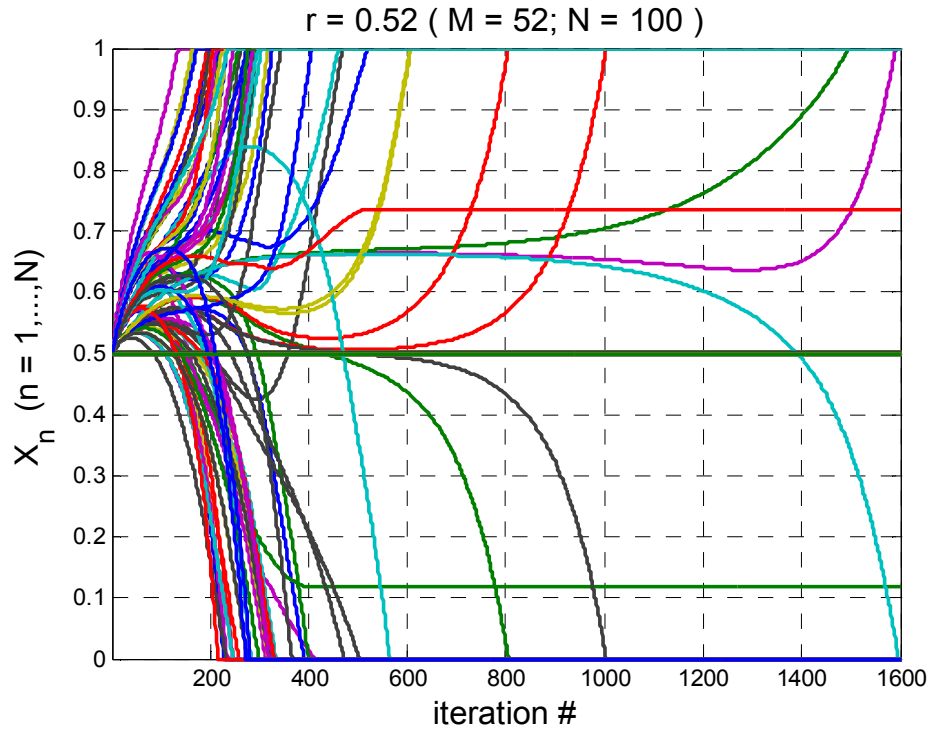 |
| 0.6 | • • • | 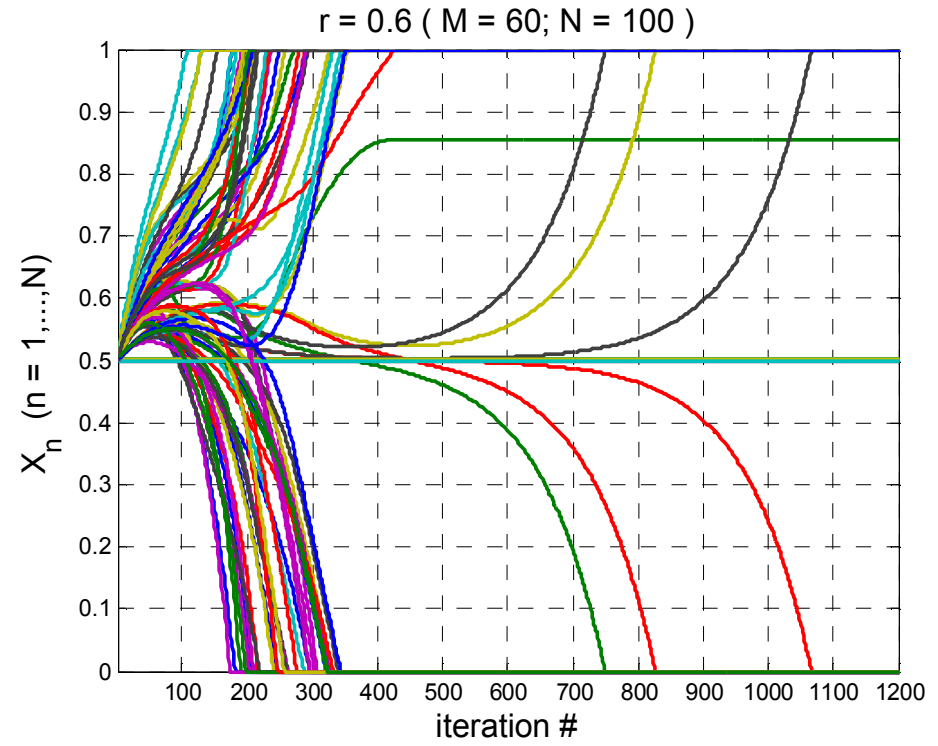 |

## 6 Discussion and Concluding Remarks

The most interesting question regarding the *BSGD* algorithm concerns its computational complexity. This is a very difficult question and we will not attempt to fully address it in this paper. Instead, some preliminary analysis of the complexity issues is given below.

Assume that the *EC3* problem at hand has one or several satisfying bit assignments. Suppose also that, without knowing that the problem is satisfiable, we have run the *BSGD* algorithm independently (using randomly selected starting points) *R* times and it has failed to find at least one satisfying solution. How confident can we be that there is a solution to be found if we continue? Let *q* be a probability to find any satisfying bit assignment in a single run of the *BSGD* algorithm in case when at least one solution does exist. Then the expected number of trials until the first success $n_s$ is $1/q$. Indeed, probability that exactly *j* trials are needed is $\Pr\{n = j\} = q(1-q)^{j-1}$ , and hence

$$n_s \equiv E[n] = \sum_{j=1}^{\infty} j \cdot \Pr\{n = j\} = q \cdot \sum_{j=1}^{\infty} j \cdot (1-q)^{j-1} = 1/q \qquad (7)$$

It is also easy to find a standard deviation $\sigma$ for the number of trials until the first success:

$$\sigma^2 \equiv E[n^2] - n_s^2 = \sum_{j=1}^{\infty} j^2 \cdot q(1-q)^{j-1} - q^{-2} = \frac{1-q}{q^2} \qquad (8)$$

Making use of the one-sided Chebyshev's inequality, we then obtain

$$\Pr\{n \ge k \cdot n_s\} \le \frac{1-q}{1-q+(k-1)^2} \ , \qquad (9)$$

where $k > 0$. If, for instance, we select $k = 11$ and run the *BSGD* algorithm $R = (11 \cdot n_s - 1)$ times without success, then, with confidence approximately 99%, we can conclude that no solutions exist and stop. The question of the complexity of the algorithm thus boils down to the estimation of $n_s$: whether it grows as a *polynomial* or *superpolynomial* of the problem size *N*. Consider a set of all points inside a unit hypercube Ω which possess the following property: when used as starting entries to the *BSGD* algorithm, they evolve to the satisfying bit assignments during its progression, i.e., they form a joint basin of attraction for all solutions. Let $\Omega^*$ and $Vol[\Omega^*]$ denote this set and its measure, respectively. Then, assuming that we choose the starting point at random with uniform distribution inside Ω, we obtain

$$n_s = q^{-1} = \frac{Vol[\Omega]}{Vol[\Omega^*]} = \frac{1}{Vol[\Omega^*]} \qquad (10)$$

Thus, we observe that the computational complexity of the *EC3* problem is linked to the properties of the corresponding harmonic cost function *F* in $\Re^N$. This connection suggests the possibility that methods of mathematical analysis, algebraic and differential topology, as well as other disciplines traditionally belonging to the domain of *continuous* *mathematics* may provide crucial insights into the most intriguing open questions in modern complexity theory.

Needless to say that the obtained results are relevant not only for the specific *EC3* problem considered throughout this paper but for *all* NP-complete problems since they transform to each other by polynomial-time reduction.

To summarize, we have demonstrated the *equivalence* of an arbitrary NP-complete problem to a problem of checking whether a level set of a specifically constructed harmonic cost function (with all diagonal entries of its Hessian matrix equal to zero) intersects with a unit hypercube in many-dimensional Euclidean space. This is the main result of the paper which can potentially lead to development of new algorithms for NP-complete problems. As an illustration of power of our method, a simple iterative algorithm (*BSGD*) belonging to the gradient descent family has been implemented for the specific NP-complete problem (*EC3* or *Positive 1-in-3 SAT*). The algorithm allows for almost linear speedup when carried out on multiple processors working in parallel. Numerical simulations confirm its good performance on problems of different sizes and reveal surprising *universal* behavior of variable flows for problems with low clauses-to-variables ratios. The computational complexity of the *BSGD* algorithm remains an open question intrinsically linked to the properties of the corresponding harmonic cost function in $\Re^N$.

## Acknowledgements

I am grateful to Dr. Mark Heiligman (IARPA) for interesting discussions.